\documentclass[aps,prb, twocolumn, showpacs, superscriptaddress]{revtex4-1}

\usepackage[dvips]{graphicx}
\usepackage{amsmath,amssymb}
\usepackage{epsfig}

\usepackage{bm}
\usepackage{graphicx}
\usepackage{color}
\usepackage{dcolumn}

\newcommand{\vk}{\mbox{$\vec k $}}

\newcommand{\IA}{IA}
\newcommand{\IAs}{IA }
\newcommand{\ISRS}{ISRS}
\newcommand{\ISRSs}{ISRS }
\begin{document}

\title{Influence of pulse width and detuning on coherent phonon generation} 

\author{Kazutaka G. Nakamura}
\email{nakamura@msl.titech.ac.jp}
\affiliation{Materials and Structures Laboratory, Tokyo Institute of Technology, 4259 Nagatsuta, Yokohama 226-8503, Japan}
\affiliation{Department of Innovative and Engineered Materials, Tokyo Institute of Technology, 4259 Nagatsuta, Yokohama 226-8503, Japan}
\affiliation{CREST, Japan Science and Technology Agency, Kawaguchi, Saitama 332-0012, Japan}

\author{Yutaka Shikano}
\email{yshikano@ims.ac.jp}
\affiliation{Materials and Structures Laboratory, Tokyo Institute of Technology, 4259 Nagatsuta, Yokohama 226-8503, Japan}
\affiliation{Research Center of Integrative Molecular Systems (CIMoS), Institute for Molecular Science, National Institutes of Natural Sciences, 38 Nishigo-Naka, Myodaiji, Okazaki, Aichi 444-8585, Japan}
\affiliation{Institute for Quantum Studies, Chapman University, 1 University Dr., Orange, CA 92866, USA}

\author{Yosuke Kayanuma}
\email{kayanuma.y.aa@m.titech.ac.jp}
\affiliation{Materials and Structures Laboratory, Tokyo Institute of Technology, 4259 Nagatsuta, Yokohama 226-8503, Japan}
\affiliation{CREST, Japan Science and Technology Agency, Kawaguchi, Saitama 332-0012, Japan}
\affiliation{Graduate School of Sciences, Osaka Prefecture University, 1-1 Gakuen-cho, Sakai, Osaka, 599-8531 Japan}

\date{\today}

\begin{abstract}
We investigated the coherent phonon generation mechanism by irradiation of an ultrashort pulse with a simple two-level model. Our derived formulation shows that both impulsive stimulated Raman scattering (\ISRS) and impulsive absorption (\IA) simultaneously occur and phonon wave packets are generated in the electronic ground and excited states by \ISRSs and \IA, respectively. We identify the dominant process from the amplitude of the phonon oscillation. For short pulse widths, the \ISRSs is very small and becomes larger as the pulse width increases. We also show that the initial phase is dependent on the pulse width and the detuning.
\end{abstract}
\pacs{78.47-J, 74.78.Bz}

\maketitle

\section{Introduction}
Coherent phonons are widely used to study phonon dynamics for a wide variety of materials such as semimetals,\cite{Cheng1991, Zeiger1992,DeCamp2001, Ishioka2006,Katsuki2013} 
semiconductors,\cite{Cho1990,Dekorsy1993,Merlin1996, Hase2003,Hu2011,Mizoguchi2013, Shimada2014, Hayashi2014} superconductors,\cite{Chwalek1991, Albrecht1991, Misochko2000, Takahashi2011} and topological insulators,\cite{Wu2008, Kamaraju2010,Norimatsu2014, Misochko2015} with pump-probe type time-resolved reflectivity measurements. An ultrashort pump pulse coherently excites optical phonons which oscillate in phase and modulate the electric susceptibility. The probe pulse monitors this modulation via a change in the reflectivity.\cite{Dekorsy2000} Thus we directly measured the time evolution of the optical phonons using the pump-probe experiment. Note that we cannot measure the time evolution using conventional frequency-domain spectroscopy because of the time resolution.

Coherent optical phonons are generated by an ultrashort pulse via photon-electron and electron-phonon coupling. The well known mechanisms for coherent phonon 
generation are impulsive stimulated Raman scattering (\ISRS)~\cite{Nelson1985} and impulsive absorption (\IA) under the displacement potential~\cite{Zeiger1992, comment} for transparent and opaque regions, respectively. However, past theoretical analyses~\cite{Kuznetsov1994, Merlin1997, Garcia2004, Shinohara2010, Shimada2012} deal with the two processes separately and show that the phonon oscillation can be fitted using sine and cosine functions for the \ISRSs and \IA, respectively. Also, several experiments showed that the phonon oscillation often shifted from the sine or cosine oscillations and its phase was dependent on the materials and phonon modes.\cite{Stevens2002, Ishioka2006, Norimatsu2014} Thus, the generation mechanism for coherent phonons is still controversial, especially for the opaque region where both the light absorption and Raman processes coexist.

The objective of this paper is to determine the generation mechanism of coherent optical phonons by evaluating the contribution of both \ISRSs and \IAs based on a simple quantum-mechanical model. We derive the time evolution of the electron-phonon coupled state by solving the time-dependent Schr\"odinger equation with density matrices using a two-level model for both processes. The quantum-mechanical calculation of the phonon dynamics in the weak coupling limit of the electron-phonon coupling makes it possible to compare the amplitudes and phases of the phonon oscillation for each process quantitatively as a function of the pulse width and the detuning. It was shown that under the resonant condition the \IAs process is dominant for coherent phonon generation and the oscillation can be described by a cosine function. When the optical pulse width increases, the contribution of the \ISRSs process increases and the phonon oscillation deviates from the cosine function. Also, the initial phase of the coherent phonons changes at a long pulse excitation because the contribution of the \ISRSs increases. As the detuning of the excitation wavelength becomes large, the amplitude of the coherent phonons in the \IAs process decreases rapidly, and the \ISRSs process becomes dominant.

\section{Model and formulation}
We consider a two-level system for the electronic state and a harmonic oscillator for the optical phonon at $\Gamma$-point ($\vk=0$). Although this is a crude model for bulk semiconductors, it describes the essential features of the generation of coherent phonons.\cite{Hayashi2014} 
It was remarked that this model can be applied to molecular vibrational spectroscopy.\cite{Pollard1990,Banin1994,Cerullo2008} The creation and annihilation operators of the LO phonon at the $\Gamma$-point with energy $\hbar\omega$ are denoted by $b^\dagger$ and $b$, respectively. It was assumed that the excited states are coupled with the LO phonon mode through the deformation potential interaction with the dimensionless coupling constant $\alpha (>0) $. We have approximated the interactions by neglecting the $\vk$-dependence of the coupling constant, and have assumed that the rigid-band shift is because of the deformation potential interaction. In the bulk crystal, the Huang-Rhys factor $\alpha^2$ is considered to be small, $\alpha^2 \ll 1$. The Hamiltonian, $H_0$, is given by
\begin{eqnarray}
H_0 &= &H_g | g \rangle \langle g | + (\epsilon + H_e )  | e \rangle \langle e | \\
H_g &= &\hbar \omega b^\dag b  \\
H_e &= &\hbar \omega b^\dag b + \alpha \hbar \omega (b^\dag +b), \label{dph}
\end{eqnarray}
where $H_g$ and $H_e$ are phonon Hamiltonians for the electronic ground $ | g \rangle$ and excited $ | e \rangle$  states, respectively. 
Using the rotating wave approximation, the interaction between the pump pulse and the electronic state is given by
\begin{eqnarray}
H_I(t) = \mu E_0 f(t) \left( e^{-i\Omega t}   | e \rangle \langle g | +  e^{i\Omega t}   
| g \rangle \langle e | \right),
\end{eqnarray}
where $\mu$ is the transition dipole moment, $\Omega$ is the central frequency of the pulse, and 
$E_0f(t)$ is the envelope of the pulse.~\cite{fcomment} It is noted that these conditions are for a Fourier-transform-limited pulse.

\begin{figure}
\begin{center}
\includegraphics[width=6.5 cm]{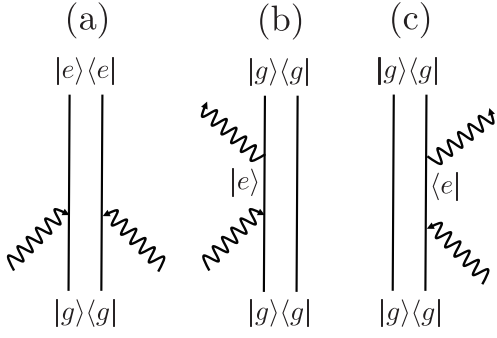}
\caption{Double-sided Feynman diagrams representing the photo-induced processes: (a) shows \IA, and (b) and (c) show \ISRS. The left and right lines represent the ket and the bra, respectively. The wavy lines represent the photons. Time runs vertically from the bottom to the top.}
\label{Feynman}
\end{center}
\end{figure}

The time evolution of the electron-phonon coupled state was obtained by solving the time-dependent Schr\"odinger equation:
\begin{equation}
i\hbar\frac{d}{dt} |\psi(t)\rangle =\{H_0 + H_I(t)\} |\psi(t)\rangle,
\end{equation}
which gives
\begin{widetext}
\begin{equation}
 | \psi (t)  \rangle = \exp\left(-  \frac{i}{\hbar } H_0 t \right)   
\exp_{+} \left( -\frac{i}{\hbar} \int_{-\infty}^t \tilde{H}_I (t^{\prime}) dt^{\prime}\right)  | \psi (-\infty)  \rangle
\end{equation}
with
\begin{equation}
 \tilde{H}_I (t^{\prime}) = \exp\left(\frac{i}{\hbar } H_0 t^{\prime} \right) H_I(t^{\prime})  \exp\left(- \frac{i}{\hbar } H_0 t^{\prime} \right) ,
\end{equation}
where $ | \psi (-\infty)  \rangle$ is the wave function for the initial state at $t=-\infty$ and 
$\exp_{+}$ is the time-ordered exponential. Here, $\tilde{H}_I (t^{\prime})$ is given by 
\begin{equation}
 \tilde{H}_I (t) = \mu E_0 f(t) \exp\left[ \frac {i}{\hbar}(\epsilon - \hbar  \Omega + H_e) t \right]
  | e \rangle \langle g | \exp\left(-\frac{i }{\hbar}H_g t\right)+ c.c..
\end{equation}
The density matrix of the electronic and phonon states is given by $\rho (t) \equiv  | \psi (t)  \rangle  \langle  \psi (t) |$. 
The $ \mu^2$ term of the density matrix $\rho ^{(2)}(t)$ is
\begin{eqnarray}
\rho ^{(2)} (t) &=&  \left(\frac{\mu E_0}{\hbar}\right)^2\exp\left(-  \frac{i}{\hbar } H_0 t \right) F(t)  
| g,0  \rangle \langle g, 0 |  F^{\dag}(t) \exp\left(  \frac{i}{\hbar } H_0 t \right)\nonumber \\
&&-\left(\frac{\mu E_0}{\hbar}\right)^2\exp\left(-  \frac{i}{\hbar } H_0 t \right)  
 G(t)   | g,0  \rangle  \langle g, 0 | \exp\left(  \frac{i}{\hbar } H_0 t \right)\nonumber \\
&&- \left(\frac{\mu E_0}{\hbar}\right)^2\exp\left(-  \frac{i}{\hbar } H_0 t \right)   | g,0  \rangle  
\langle g, 0 |   G^{\dag}(t) \exp\left(  \frac{i}{\hbar } H_0 t \right),
\label{density}
\end{eqnarray}
where the initial state is the electronic ground state and the zero phonon state $| g,0  \rangle$, 
and $F(t)$ and $ G(t)$ are given by
\begin{eqnarray}
F(t) &=& \int_{-\infty}^t dt^{\prime} f(t^{\prime}) A^{\dag} (t^{\prime})B(t^{\prime}) | e \rangle \langle g |, \\ 
G(t) &=&  \int_{-\infty}^{t} dt^{\prime}\int_{-\infty}^{t^{\prime}}  dt^{\prime\prime} f(t^{\prime})   f(t^{\prime\prime}) B^{\dag} (t^{\prime}) A (t^{\prime}-t^{\prime\prime})
B(t^{\prime\prime})  | g \rangle \langle g | .
\end{eqnarray}
\end{widetext}
Both $A(t)$ and $B(t)$ are phonon propagators given by 
\begin{eqnarray}
A(t) &=& \exp\left[ -\frac{i}{\hbar} (\epsilon - \hbar  \Omega + H_e) t \right] ,\\
B(t) &=& \exp\left( -\frac{i}{\hbar}  H_g t\right).
\end{eqnarray}
Each term in the right hand side of Eq. (\ref{density}) corresponds to the process described by 
the double-sided Feynman diagrams (Fig. \ref{Feynman} (a -- c)).
The first term corresponds to \IAs (Fig. \ref{Feynman} (a)) and the second and third terms correspond to the \ISRS. Hereafter, each term on the right side of Eq. (\ref{density}) for 
$\rho ^{(2)} (t) $ is abbreviated to $\rho ^{(2)}_a (t) $, $\rho ^{(2)}_b (t) $, and $\rho ^{(2)}_c (t) $, 
respectively. It should be emphasized that these three processes occur equivalently in quantum 
mechanics but have different signs in three terms in Eq. (\ref{density}).

The density matrix $\rho^{(2)}_a (t)$ has only the component $ \langle e | \rho^{(2)}_a (t) | e \rangle$, 
which corresponds to the \IAs process, where the phonon oscillates on the adiabatic potential of 
the excited state after optical absorption. However, the density matrix $\rho^{(2)}_b (t)$ has 
only a nonzero matrix element, $ \langle g| \rho^{(2)}_b (t) | g \rangle$, and the coherent phonons 
are generated in the electronic ground state. The density matrix $ \rho^{(2)}_c (t) $ is the Hermitian conjugate to $ \rho^{(2)}_b (t)$. 
These two processes represent the \ISRS, in which the phonon oscillates on the ground state adiabatic potential. 
The coherent phonon dynamics can be investigated by calculating the mean value of the phonon coordinate $\langle Q (t) \rangle = \mathrm{Tr}\{ Q \rho ^{(2)} (t) \}$, 
where $Q \equiv \sqrt{\hbar/2\omega}\left(b+b^\dag\right)$ and $\mathrm{Tr}$ indicates that the trace should be taken over the electronic and phonon variables. 
The displacement of each process is separated as follows: 
\begin{equation}
\langle Q(t)\rangle =\langle Q_A(t) \rangle +\langle Q_R(t) \rangle,
\end{equation}
where
\begin{equation}
\langle Q_A(t) \rangle=\mathrm{Tr} \{Q \rho ^{(2)}_a (t) \} ,
\end{equation}
and
\begin{equation}
\langle Q_R(t) \rangle=\mathrm{Tr} \{Q \rho ^{(2)}_b (t) \} + \mathrm{Tr} \{Q\rho ^{(2)}_c (t)\}.
\end{equation}

The expressions for the density matrices are easily calculated in our model. For the \IAs process, 
we find 
\begin{equation}
\langle e| \rho^{(2)}_a (t) | e \rangle =\left(\frac{\mu E_0}{\hbar}\right)^2 |\varphi (t)\rangle
\langle \varphi (t) |,
\end{equation}
where $|\varphi(t) \rangle $ is the phonon wave function given by
\begin{widetext}
\begin{eqnarray}
|\varphi(t) \rangle &=& \int_{-\infty}^t dt^{\prime} f(t^{\prime}) \exp\left[ -\frac{i}{\hbar}
\left(\epsilon -\hbar\Omega+H_e\right)(t-t^{\prime})\right] |0\rangle \nonumber\\
&=& \int_{-\infty}^t dt^{\prime} f(t^{\prime})  U^\dag (\alpha) |\alpha e^{-i\omega(t-t^{\prime})}\rangle 
e^{-\frac{i}{\hbar}(\epsilon -\hbar\Omega-\alpha^2 \hbar\omega )
(t-t^{\prime})}.
\end{eqnarray}
Here, the shift operator $U(\beta)$ is defined for an 
arbitrary complex parameter $\beta$ by 
$U(\beta) \equiv \exp[\beta b^\dag -\beta^* b]$ and the coherent state $|\beta \rangle$ is given by $|\beta\rangle =U(\beta)|0\rangle $.
Then, we obtain
\begin{eqnarray}
\langle Q_A(t)\rangle &=& \alpha\left(\frac{\mu E_0}{\hbar}\right)^2  \sqrt{\frac{\hbar}{2\omega}}
\int_{-\infty}^t dt^{\prime}\int_{-\infty}^t  dt^{\prime\prime} f(t^{\prime})f(t^{\prime\prime}) \left(e^{-i\omega(t-t^{\prime})}+ e^{i\omega(t-t^{\prime\prime})}-2\right) \nonumber \\
& & \times\exp\left[- \alpha^2 \{1+i\omega (t^{\prime}-t^{\prime\prime})-e^{i\omega(t^{\prime}-t^{\prime\prime})}\}\right] e^{i(\epsilon -\hbar\Omega)(t^{\prime}-t^{\prime\prime})/\hbar}.
\label{QAgeneral}
\end{eqnarray}

For the \ISRS, we can write
\begin{equation}
\langle g| \rho^{(2)}_b (t) | g \rangle =-\left(\frac{\mu E_0}{\hbar}\right)^2 |\chi(t)\rangle
\langle 0 |,
\end{equation}
where $|\chi(t) \rangle $ is given by
\begin{eqnarray}
|\chi(t) \rangle &=& \int_{-\infty}^t dt^{\prime}\int_{-\infty}^{t^{\prime} } dt^{\prime\prime} f(t^{\prime}) f(t^{\prime\prime}) 
\exp\left[ -\frac{i}{\hbar}H_g(t-t^{\prime})\right]
\exp\left[ -\frac{i}{\hbar}
\left(\epsilon -\hbar\Omega+H_e\right)(t^{\prime}-t^{\prime\prime})\right] |0\rangle \nonumber\\
&=& \int_{-\infty}^t dt^{\prime} \int_{-\infty}^{t^{\prime} }dt^{\prime\prime} f(t^{\prime}) f(t^{\prime\prime}) U^\dag (\alpha e^{-i\omega (t-t^{\prime})})
|\alpha e^{-i\omega(t-t^{\prime\prime})}\rangle e^{-i(\epsilon -\Omega -\alpha^2
\hbar\omega)(t^{\prime}-t^{\prime\prime})/\hbar}.
\end{eqnarray}
The mean value of $Q_R(t)$ for the \ISRSs process is then given by
\begin{eqnarray}
\langle Q_R(t) \rangle &=& \alpha \left(\frac{\mu E_0}{\hbar}\right)^2  
\sqrt{\frac{\hbar}{2\omega}}\int_{-\infty}^t dt^{\prime} \int_{-\infty}^{t^{\prime} }dt^{\prime\prime} f(t^{\prime}) f(t^{\prime\prime}) 
\left(e^{-i\omega (t-t^{\prime})}-e^{-i\omega (t-t^{\prime\prime})}\right) \nonumber\\
&&\times \exp\left[-\alpha^2\{1-i\omega(t^{\prime}-t^{\prime\prime})-e^{-i\omega(t^{\prime}-t^{\prime\prime})}\}\right]
e^{-i(\epsilon - \hbar\Omega)(t^{\prime}-t^{\prime\prime})/\hbar}+ c.c..
\label{QRgeneral}
\end{eqnarray}
\end{widetext}
Formulae (\ref{QAgeneral}) and (\ref{QRgeneral}) are general expressions for the expectation values of the phonon coordinate in \IAs and \ISRS, respectively. 

\section{Results and Discussion}
Let us assume that the pulse-envelope function $f(t)$ is a real quantity 
localized in the region $|t|\lesssim\sigma$ around $t=0$ with the normalization condition 
$\int_{-\infty}^\infty f(t) dt  =1$. As a typical example, we set the Gaussian function to be 
\begin{equation}
f(t)=\frac{1}{\sqrt{\pi}\sigma}\exp(-t^2/\sigma^2).
\label{Gauss}
\end{equation}
In what follows, $(\mu E_0/\hbar)^2 \equiv 1$.
\subsection{Resonant condition}
\begin{figure}[t]
\centering
\includegraphics[width=4.5cm]{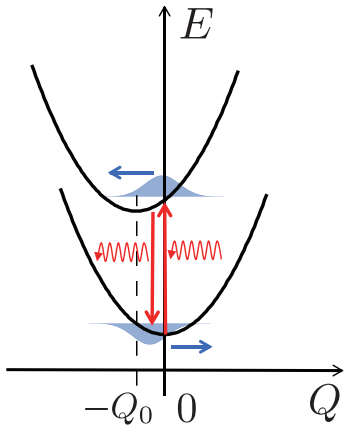}
\caption{Schematic illustration of the \IAs and \ISRSs processes on the adiabatic potential curves for the electronic ground and excited states. The horizontal small arrows indicate the direction of the initial motion of the phonon in the electronic excited and ground states.}
\label{APES}
\end{figure}
In this subsection, we focus our attention to the case of resonant excitation with weak 
electron-phonon coupling. Setting $\epsilon-\hbar\Omega=0$ and neglecting the terms 
of $\alpha^2$ order, we simplify the expressions to give
\begin{eqnarray}
\langle Q_A(t) \rangle &=&\alpha   
\sqrt{\frac{\hbar}{2\omega}}\int_{-\infty}^t dt^{\prime} \int_{-\infty}^{t }dt^{\prime \prime}f(t^{\prime}) f(t^{\prime \prime}) \nonumber\\
&&\times\left[ \cos\omega(t-t^{\prime})+\cos\omega(t-t^{\prime \prime})-2 \right],
\label{generalQA}
\end{eqnarray}
\begin{eqnarray}
\langle Q_R(t)\rangle  
&=&2\alpha  
\sqrt{\frac{\hbar}{2\omega}}\int_{-\infty}^t dt^{\prime} \int_{-\infty}^{t^{\prime}}dt^{\prime \prime}f(t^{\prime}) f(t^{\prime \prime}) \nonumber\\
&&\times
\left[ \cos\omega(t-t^{\prime}) -\cos\omega(t-t^{\prime \prime}) \right].
\label{generalQR}
\end{eqnarray}

We discuss the general features of $\langle Q_A(t)\rangle$ and $\langle Q_R \rangle$ based on the formulae (\ref{generalQA}) and (\ref{generalQR}). 
First let us discuss the behavior of $\langle Q_A(t) \rangle $. It is obvious from Eq.(\ref{generalQA}) that $Q_A(t)\leq 0$ since the phonon oscillates on the 
excited state potential energy curve as shown schematically in Fig. \ref{APES}. 
In the case of the Gaussian pulse (\ref{Gauss}), and for $t \gg \sigma$, the integral is approximately carried out by extending the upper limit to $\infty$, and we find the cosine-like oscillation around 
the new equilibrium point $ -Q_0 (\equiv -2\alpha\sqrt{\hbar/2\omega} )$ in the excited state, 
\begin{equation}
\langle Q_A(t)\rangle =Q_0 \left(e^{-\sigma^2\omega^2/4}\cos \omega t - 1\right).
\label{Qanalytic}
\end{equation}
In the short pulse limit $\sigma\omega\rightarrow 0$, this reduces to 
\begin{equation}
\langle Q_A(t)\rangle =Q_0 \left(\cos \omega t - 1\right),
\end{equation}
which is in agreement with the coherent state. In the long pulse limit,
$\sigma\omega \gg 1$, however, the oscillation disappears and $\langle Q_A(t)\rangle$ changes gradually during the pulse duration from $0$ to the new equilibrium of the lowest vibrational state $\langle Q_0\rangle$ in the excited state. Thus, Eq. (\ref{Qanalytic}) clearly shows the changeover of the phonon dynamics induced by the electronic excitation, from the sudden transition limit $\sigma\omega\rightarrow 0$ to the adiabatic change limit $\sigma\omega\rightarrow \infty$.

In the \ISRSs process described by $\langle Q_R(t)\rangle$, the oscillation 
is induced by the sudden occurrence of momentum because of the excitation and the deexcitation 
in the impulsive stimulated Raman process. Therefore, its motion is sine-like. In the case of the resonant excitation, $\langle Q_R(t)\rangle$ is calculated as follows. 

For the time $t$ after the passage of the optical pulse, $t \gg \sigma$, the time-ordered integral in Eq. (\ref{generalQR}) can be evaluated by extending the upper limit of the integral to infinity: 
\begin{widetext}
\begin{equation}
\langle Q_R(t)\rangle=Q_0 e^{-i\omega t}\int_{-\infty}^\infty dt^{\prime} \int_{-\infty}^{t^{\prime} }dt^{\prime \prime} f(t^{\prime})f(t^{\prime \prime})
\left(e^{i\omega t^{\prime}} - e^{i\omega t^{\prime \prime}}\right) + c.c..
\end{equation}
Using new variables $s\equiv(t^{\prime}+t^{\prime \prime})/2$ and  $u\equiv t^{\prime}-t^{\prime \prime}$, we find
\begin{equation}
\langle Q_R(t)\rangle = Q_0 \frac{2 i e^{-i\omega t}}{\pi\sigma^2} \int_{-\infty}^{\infty} ds \int_0^\infty du
\exp\left[ -\frac{2}{\sigma^2}s^2-\frac{1}{2\sigma^2}u^2\right] e^{i\omega s} \sin \left[ \frac{\omega}{2}u \right] +c.c. 
= A\sin\omega t.
\label{sine}
\end{equation}
\end{widetext}
The useful formulae~\cite{comment2} are included with the coefficient $A$ given by 
\begin{equation}
A \equiv Q_0 \frac{4}{\sqrt{\pi}}e^{-\sigma^2\omega^2/4}\int_0^{\sigma\omega/2\sqrt{2}} e^{t^2} dt.
\end{equation}
Equation (\ref{sine}) tells us that the phonon oscillation is induced by the impulsive generation of momentum because the excitation and the de-excitation occurs through \ISRS. Therefore, its motion is sine-like. This also indicates that the amplitude of the oscillation $A$ takes a maximum value at an intermediate value of the pulse width $\sigma$ because $A\rightarrow 0$ both in the limits $\sigma\rightarrow 0$ ($\delta$-function pulse) and $\sigma\rightarrow \infty$. In other words, it needs a finite duration in the electronic excited state for the phonon to get momentum. Since the coefficient $A$ must be positive, the phonon wave packet begins to move in the direction opposite to $\langle Q_A (t)\rangle$. This counterintuitive phenomenon can be understood as follows. 
In the process of \ISRSs described by Fig. \ref{Feynman} (b) and (c), the phonon wave function in the excited state gives a contribution that is the negative of the ground state. This is explicitly shown by the negative signs of the $\rho^{(2)}_b (t)$ and $\rho^{(2)}_c (t)$ in Eq. (\ref{density}). In other words, a \textit{hole} is created in the ground state wave function 
of the phonon\cite{Banin1994,Kayanuma2000} by \ISRSs as shown in Fig. \ref{APES} so that the expectation value $\langle Q_A(t)\rangle$ moves in the opposite direction to that in the \IAs process. 
\begin{figure}
\centering
\includegraphics[width=7cm]{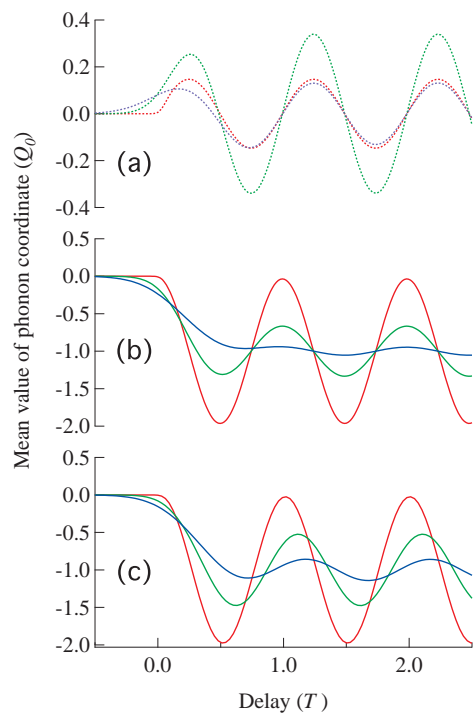}
\caption{(Color online) Time evolution of the mean value of the atomic displacement with pulse width (full width at half maximum) of $0.1 T$ (red), $0.55 T$ (green), and $0.9 T$ (blue) for 
the vibrational period ($T$) of the phonon: (a) the dotted curves represent $\langle Q_R (t) \rangle$; (b) the solid curves represent $\langle Q_A (t) \rangle$; (c) the solid curves represent $\langle Q (t) \rangle$.}
\label{Q_cal}
\end{figure}

In Figs. \ref{Q_cal} (a) and (b), the numerical results of the dependence of the pulse-width on $\langle Q_A(t)\rangle$ (solid lines) and $\langle Q_R(t)\rangle$ (dashed lines) are shown as a function of the delay-time for the Gaussian pulses. The vibrational period of the coherent phonons was set to be \textit{T}.\cite{comment3} For $\langle Q_A(t)\rangle$, the approximate formula (\ref{Qanalytic}), agrees with the exact results in the region of time after the passage of the pulse. Violent oscillations of the coherent phonons changes to a gradual adaptation of a new equilibrium as the pulse width becomes large. 

However, for $\langle Q_R(t)\rangle$, there is an optimum value of the pulse width that maximizes the amplitude of the oscillation. To discriminate experimentally between $\langle Q_A(t)\rangle$ and $\langle Q_R(t)\rangle$, it is necessary to use the electronic state-selective measurement of the coherent phonons. However, in the commonly used transient reflectivity or transmissivity measurements, the electronic states are not identified. One can get information only on the total value $\langle Q(t) \rangle\equiv \langle Q_A(t)\rangle + \langle Q_R(t)\rangle$, which is also shown in Fig. \ref{Q_cal} (c). 
$\langle Q(t) \rangle$ shows an oscillation with the frequency $ \omega$, which is almost cosine-like at short pulse widths such as $0.1 T$. As the pulse width increases, the contribution from the \ISRSs process increases and the initial phase of the oscillation changes.

\begin{figure}
\centering
\includegraphics[width=6cm]{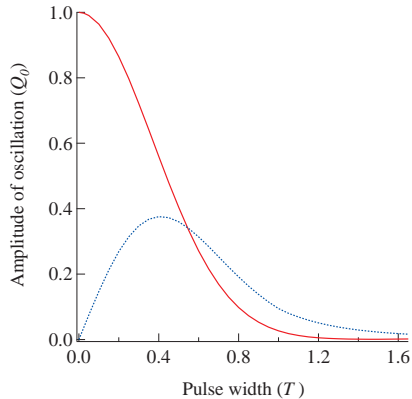}
\caption{(Color online) The dependence of the pulse-width on the amplitude of the oscillation part of the phonons, for $\langle Q_A (t) \rangle$ (red) and $\langle Q_R (t) \rangle$ (blue). \textit{T} is the vibrational period of the phonons.}
\label{amplitude}
\end{figure}
In Fig. \ref{amplitude}, the amplitudes of the oscillation are plotted for \IAs (red line) and \ISRSs (blue line) as a function of the pulse-width (FWHM). The dominant process changes from \IAs to \ISRSs around the pulse width of $0.55 T$. This phenomenon may be experimentally observed from the initial phase of the phonon oscillation by experiments that precisely control the pulse width and frequency chirping.
\subsection{Detuning effect}
The detuning from the resonant condition was studied for $\Delta E \equiv \epsilon - \hbar \Omega > 0$ 
by evaluating Eqs. (\ref{QAgeneral}) and (\ref{QRgeneral}). 
It is worth noting that the $\langle Q_A(t)\rangle$ is also described analytically for $t \gg \sigma$ by 
\begin{equation}
\langle Q_A(t)\rangle =Q_0 e^{-\sigma^2 \xi^2/2}  \left(e^{-\sigma^2 (\omega^2+2 \xi \omega )/4}\cos \omega t - 1\right)
\label{Qanalytic2}
\end{equation}
with $\xi \equiv (\epsilon -\hbar \Omega)/\hbar$.

However, $\langle Q_R (t)\rangle$ can be calculated using 
\begin{widetext}
\begin{equation}
\langle Q_R (t) \rangle = \frac{1}{\sqrt{2\pi}} e^{-\sigma^2\omega^2/8} e^{-i\omega t} 
\int_0^\infty du e^{-u^2/2\sigma^2} \left( e^{i\omega u/2} -e^{-i\omega u/2}\right) e^{-i \xi u} + c.c..
\end{equation}
For general values of the detuning, $\xi$, the above integral yields a complex value, so that the phase of the oscillation of $\langle Q_R (t) \rangle$ changes gradually as the detuning becomes large. We evaluated the initial phase in the limit of large detuning $\xi \sigma \gg 1$ using 
\begin{equation}
\int_0^\infty du e^{-u^2/2\sigma^2-i \eta u} = \sqrt{\frac{\pi}{2}}\sigma e^{-\sigma^2 \eta^2/2}-
i\sqrt{2}\sigma e^{-\sigma^2 \eta^2/2}\int_0^{\sigma \eta/\sqrt{2}}e^{t^2} dt
\end{equation}
\end{widetext}
with the positive constant $\eta$.
It is obvious that the real part becomes negligible when compared with the imaginary part in the limit $\sigma \eta \gg 1$. Therefore, in the case of large detuning, we can neglect the real part and obtain 
\begin{equation}
\langle Q_R (t)\rangle =B\sin\omega t,
\label{sine2}
\end{equation}
where
\begin{equation}
B \equiv Q_0\frac{2}{\sqrt{\pi}} e^{-\sigma^2\omega^2/8} \left\{D\left(\xi +\frac{\omega}{2}\right)
-D\left(\xi -\frac{\omega}{2}\right)\right\},
\end{equation}
and 
\begin{equation}
D(x)\equiv e^{-\sigma^2 x^2/2}\int_0^{\sigma x/\sqrt{2}} e^{t^2} dt.
\end{equation}

In the limit of large detuning, the asymptotic form of $D(x)$ can be obtained from the inverse power-series expansion: 
\begin{equation}
D(x)\sim a_1 x^{-1}+a_2 x^{-2} + \cdots.
\end{equation}
Inserting the above expression into the differential equation, $dD(x)/dx=-\sigma^2 x D(x)+\sigma/\sqrt{2}$, 
we find to the lowest order term
\begin{equation}
D(x) \simeq \frac{1}{\sqrt{2}\sigma x},
\end{equation}
and 
\begin{equation}
B\simeq - Q_0 \sqrt{\frac{2}{\pi}} e^{-\sigma^2\omega^2/8}\frac{\omega}{\sigma \xi^2},
\label{BforlargeD}
\end{equation}
where we have used the approximation
\begin{equation}
D \left( \xi+\frac{\omega}{2} \right) - D \left( \xi-\frac{\omega}{2} \right) \simeq 
\frac{dD}{d\xi}\omega.
\end{equation}

\begin{figure}
\centering
\includegraphics[width=7cm]{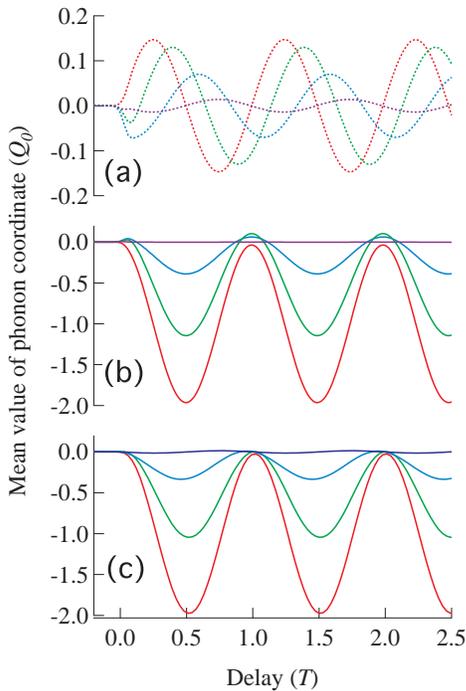}
\caption{(Color online) The time evolution of the mean value of the phonon coordinate with detuning $\Delta E$ of $0 \hbar \omega$ (red: no detuning), $3 \hbar \omega$  (green), 
$5 \hbar \omega$  (blue), and $10 \hbar \omega$ (purple). (a) The dotted curves represent $\langle Q_R (t) \rangle$, (b) the solid curves represent $\langle Q_A (t) \rangle$ and (c) the solid curves represent $\langle Q (t) \rangle$. $T$ is the vibrational period of the phonons. The pulse width was set to $0.1 T$.}
\label{Q_cal2}
\end{figure}
Therefore, we find $B<0$ in the limit of large detuning. The initial phase of the sine-like oscillation in \ISRSs changes by $\pi$ as the detuning changes from $0$ to large values. 
Equation (\ref{BforlargeD}) is the general form for the dependence of the amplitude of the coherent phonon oscillation on the detuning $\xi$ and pulse-width $\sigma$ in the case of pumping in the transparent region.

\begin{figure}
\centering
\includegraphics[width=6cm]{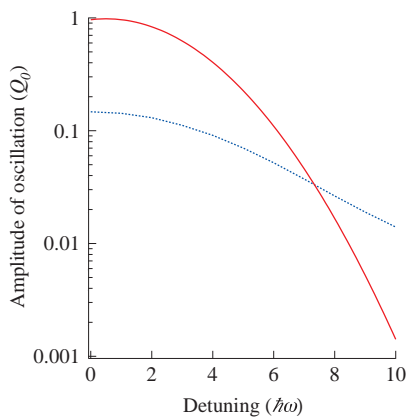}
\caption{(Color online) The detuning dependence of the amplitude of the oscillation of the phonons, for $\langle Q_A (t) \rangle$ (red) and $\langle Q_R (t) \rangle$ (blue). The pulse width was set to $0.1 T$.}
\label{amplitude2}
\end{figure}

Figure \ref{Q_cal2} shows the time evolution of the atomic displacements $\langle Q_A(t)\rangle$, $\langle Q_R(t)\rangle$, 
and $\langle Q(t)\rangle$ with a pulse width of $0.1 T$. The center position of the atom around which $\langle Q_A(t)\rangle$ oscillates 
approaches zero as $\Delta E$ increases. $\langle Q_R(t)\rangle$ starts to move in the same direction as $\langle Q_A(t)\rangle$ at large detuning, as alluded to before.
Figure \ref{amplitude2} shows the amplitude of the oscillation of the phonons. The oscillation amplitude of both $\langle Q_A(t)\rangle$ and $\langle Q_R(t)\rangle$ decreases as $\Delta E$ increases. For large detuning, the oscillation amplitude of $\langle Q_R(t)\rangle$ becomes larger than that of the $\langle Q_A(t)\rangle$ since no light absorption occurs, and 
the dominant coherent phonon generation process is subject to the \ISRSs mechanism. 
Both of the numerical values of the oscillation amplitude for $\langle Q_A(t)\rangle$ and $\langle Q_R(t)\rangle$ agree with the analytical formulas, namely Eq. (\ref{Qanalytic2}) for $\langle Q_A(t)\rangle$ and Eq. (\ref{sine}) for $\langle Q_R(t)\rangle$ where $\xi=0$ and Eq. (\ref{sine2}) where $\xi \gg \omega$.

\section{Conclusion}
In this work, we investigated the generation mechanisms of coherent optical phonons using a simplified two-level model with resonant excitation conditions. The quantum-mechanical calculations indicate that both the optical phonons in the electronic excited $| e\rangle $ and ground $| g\rangle $ states are excited via impulsive absorption and stimulated Raman scattering by irradiation of the femtosecond pulse at resonance. In the short pulse limit, only the optical phonon in the excited state is driven to the coherent state, but the phonons are not excited in the electronic ground state. As the pulse width increased, the amplitude of $| e\rangle $ decreased while that of $| g\rangle $ increased. The mean value of the atomic displacement, $\langle Q_A \rangle$ and $\langle Q_R \rangle $, started to move in opposite directions in $| e\rangle $ and $| g\rangle $, respectively. In the long pulse limit, both of the amplitudes of $\langle Q_A \rangle$ and $\langle Q_R \rangle $ tended to zero. In the intermediate conditions, one should 
consider the superposition of \IAs and \ISRSs processes, although \IAs is generally dominant. Our proposed model shows that both well-known coherent phonon mechanisms, \IAs and \ISRS, occur depending on the pulse length and the detuning. The initial phase of the phonon oscillation in the case of resonant excitation changed depending on the pump pulse-width because of competition between the \IAs and \ISRSs mechanisms. This indicates that care must be taken in making generation mechanism arguments based on the initial phase.

In the present work, only the symmetric mode of the phonons was considered. The model can be 
easily extended to also include asymmetric Raman modes, which rotate the polarization 
of the incident photons.\cite{JJAP} Also, in the usual optical detection of coherent phonons, 
atomic displacements are detected indirectly through the modulation of the electric susceptibility. 
The extension of the present study to include these processes will be shown in a forthcoming paper. 
Further, the present work investigates the coherent control of the lattice vibrations by a pair of optical pulses 
with a well-defined phase.~\cite{Katsuki2013,Hayashi2014,Castella1999} The extension of the present study 
can be also applied to such double-pulse system. 
Finally, the environmental effects on the electronic and the phonon (vibrational) states is to be considered.\cite{Mukamel1995} 
It is important to analyze these effects to determine the lifetime and the decoherence mechanism of the coherent phonons.
\begin{acknowledgements}
The authors thank K. Goto of Tokyo Institute of Technology for his effort in the early stage of this work. 
This work was partially supported by Core Research for Evolutional Science and Technology (CREST) of the Japan Science and Technology Agency (JST), 
JSPS KAKENHI Grant Number 25400330 and 15K13377, the Collaborative Research Project of Materials and Structures Laboratory, NINS Youth Collaborative Project, 
and the Joint Studies Program of the Institute for Molecular Science.
\end{acknowledgements}


\begin{thebibliography}{99}
\bibitem{comment} This is often called a displacive excitation of coherent phonon (DECP) mechanism.
\bibitem{fcomment} It is emphasized that there is no direct coupling between photon and phonon. Throughout the deformation potential in Eq. (\ref{dph}), which represents the dielectric polarization, 
the phonon state can be modulated.
\bibitem{comment3} The typical value of  the vibrational period is a few hundreds of femtoseconds for the optical phonons: for example, $333$ fs for $A_{1g}$ 
of bismuth and $115$ fs for longitudinal optical phonons of gallium arsenide.
\end{thebibliography}
\end{document}